\documentclass{iopart}
\usepackage{iopams}
\usepackage{amsopn}
\usepackage{cite}
\usepackage{mathrsfs}
\usepackage{graphicx}
\usepackage{color}
\ifx\pdfoutput\undefined
\providecommand{\eprint}[2][]{{\texttt{#2}}}%
\providecommand{\ejref}[2][]{{#2}}%
\providecommand\ead[1]{\vspace*{5pt}\address{E-mail: \mailto{#1}}}
\def\mailto#1{{\texttt{\MakeLowercase{#1}}}}
\else
\usepackage{hyperref}
\hypersetup{%
  pdftitle={Self-Dual Bending Theory for Vesicles},
  pdfauthor={J\'er\^ome~Benoit, Elizabeth~von~Hauff, and Avadh~Saxena},
	pdfsubject={PACS: 02.40.-k, 87.16.Dg, 46.70.Hg, 05.45.Yv; AMS: 53C80, 74K15},
	pdfkeywords={bending elasticity; Bogomol'nyi decomposition; sine-Gordon kink; vesicle},
  pdfpagemode=UseOutlines,
  pdfstartpage=1,
  pdfhighlight=/O,
  pdfview=FitH,
  pdfstartview=FitH,
  urlcolor=blue,
  }
\IfFileExists{thumbpdf.sty}{\usepackage{thumbpdf}}{}
\providecommand{\eprint}[2][]{\href{http://xxx.lanl.gov/abs/#2}{\texttt{#2}}}%
\providecommand{\ejref}[2][]{\href{http://stacks.iop.org/#1}{#2}}%
\providecommand\ead[1]{\vspace*{5pt}\address{E-mail: \mailto{#1}}}
\def\mailto#1{\href{mailto:#1}{\texttt{\MakeLowercase{#1}}}}
\fi

\input{\jobname.rty}

\begin{document}
\title[Self-Dual Bending Theory for Vesicles]{%
    Self-Dual Bending Theory for Vesicles%
		$^\ast$%
  }
\author{%
  J\'er\^ome~Benoit$^1$,
	Elizabeth~von~Hauff$^2$\
  and Avadh~Saxena$^3$%
  }
\address{$^\ast$\
  \ejref[0951-7715/17/57]{\textsc{Nonlinearity} \textbf{17} (2004) 57--66}
  [\eprint{cond-mat/0210441}]%
  }%
\address{$^1$\
	Physics Department,
	University of Crete and Foundation for Research and Technology-Hellas,
	PO Box 2208,
	GR-71003 Heraklion, Crete,
	Greece%
	}
\address{$^2$\
	Faculty of Physics,
	Department of Energy and Semiconductor Research,
	University of Oldenburg,
	D-26111 Oldenburg,
	Germany%
	}
\address{$^3$\
	Theoretical Division,
	Los Alamos National Laboratory,
	Los Alamos,
	NM 87545,
	USA%
	}
\ead{jgmbenoit@wanadoo.fr}
\date{\texttt{cond-mat}/0210441}
\begin{abstract}
We present a self-dual bending theory
that may enable a better understanding of highly nonlinear 
global behavior observed in biological vesicles.
Adopting this topological approach for spherical vesicles of revolution
allows us to describe them as frustrated sine-Gordon kinks.
Finally, to illustrate an application of our results,
we consider a spherical vesicle globally distorted by two polar latex beads.
\end{abstract}

\pacs{02.40.-k, 87.16.Dg, 46.70.Hg, 05.45.Yv}
\ams{53C80, 74K15}

\section{Introduction} 
Our primary motivation is to contribute to the understanding
of highly nonlinear global behavior observed in vesicles.
Such cases include closed vesicles with low genus
exhibiting spontaneous conformal transformation
\cite{PLIPO,OSSCDGV,VTTOMST},
spherical vesicles globally distorted by a single latex bead
\cite{Koltover1999},
and spherical vesicles with several arms
\cite{SFV}.
Whereas the difference in scale between
the vesicle's membrane thickness
and its overall size
allows for the vesicle to be described
as an embedded surface,
the understanding of vesicle morphology is currently founded
on the concept of bending elasticity
\cite{XHBP8,XAmMn,CFMV,Canham,HelfrichZN1,HelfrichZN2}.
To investigate vesicles, biophysicists widely invoke
a harmonic bending Hamiltonian,
historically introduced by Helfrich \cite{HelfrichZN1},
and perform polynomial approximations
to describe matter interactions.
However,  
within the context of exploring nonlinear global behaviors
such an approach may appear inappropriate
to a geometric topologist.
In this paper
an attempt to construct
a more suitable covariant field theory
for the bending behavior of vesicles,
outlined in a previous work \cite{BDVAG},
is illustrated through
spherical vesicles of revolution.

Customarily,
the Monge representation
(surface equations)
of the vesicle shape \cite{XAmMn,CFMV}
leads to a local characterization of the surface
by its two principal curvatures \cite{Struik,MDGCSM,GPI},
and thereby to an expression for the bending energy
as an expansion in the principal curvatures
up to a given order and with respect to some desired symmetries:
\newcounter{counterBEDExpansion}\setcounter{counterBEDExpansion}{1}%
(\roman{counterBEDExpansion})\stepcounter{counterBEDExpansion}~%
the even symmetric expansion up to second order
corresponds to the curvature energy density introduced by Canham
\cite{Canham}
which has one phenomenological parameter;
(\roman{counterBEDExpansion})\stepcounter{counterBEDExpansion}~%
the most general symmetric expansion up to second order
is the well known Helfrich curvature energy density
which contains three phenomenological parameters
\cite{XHBP8,XAmMn,CFMV,HelfrichZN2};
(\roman{counterBEDExpansion})\stepcounter{counterBEDExpansion}~%
the antisymmetric expansion up to second order
fits the deviatoric bending contribution density suggested by Fischer
\cite{BSLBIII,BSLBV}
which has two phenomenological parameters;
(\roman{counterBEDExpansion})\stepcounter{counterBEDExpansion}~%
expansion of higher order can be envisaged likewise
\cite{GoetzHelfrich1996}.
By assumption,
the general symmetric expansions in the principal curvatures
can be formulated as expansions in the mean curvature
(mean of the principal curvatures)
and in the Gaussian curvature
(product of the principal curvatures):
the harmonic Helfrich curvature energy density is quadratic
in the mean curvature and linear in the Gaussian curvature
\cite{XHBP8,XAmMn,CFMV,HelfrichZN2}.
Commonly,
as long as the vesicle remains in the same topological class,
the Gaussian curvature term is dropped out
\cite{XHBP8,XAmMn,CFMV}
by invoking the Gauss-Bonnet theorem \cite{Struik,MDGCSM,GPI} which 
claims that the total integral of the Gaussian curvature over the 
vesicle surface depends only on the topology of the shape%
---the total Gaussian curvature measures the genus of the shape.
From the perspective of a geometric topologist,
ignoring such a  global scale property
may be undesirable and may essentially favor
the notion of bending elasticity on the local scale.
Although this local approach
allows a large variety of phenomena to be understood
\cite{SFV,XHBP8,XAmMn,CFMV,CDCDV,MVHTG},
it may certainly preclude a deep understanding of some global bending phenomena
as it fails to reveal the topological classification of vesicles.

In contrast,
the fundamental theorem of surface theory
\cite{Struik,MDGCSM,GPI,BishopCrittenden}
leads to a description of the surface
by a prescribed metric tensor and a prescribed shape tensor
related to each other by integrability conditions.
We should therefore inquire
how bending elasticity may be formulated
within a covariant field theory for vesicles,
and ultimately determine  
which principle may dictate bending elasticity. 
With this in mind,
the observation of closed vesicles with low genus
exhibiting spontaneous conformal transformation
\cite{PLIPO,OSSCDGV,VTTOMST}
strongly suggests
imposing a covariant functional 
which reveals the underlying topology
and which is globally invariant
under smooth conformal transformations as a bending Hamiltonian.
In general,
symmetries and conservation laws
are connected by the N{\oe}ther theorem
\cite{NOETHER,NOETHERTAVEL,Ryder}.
Nevertheless,
in this context,
the conserved entity emerges from topology
while the pertinent transformations are restricted by topology:
two vesicles of different genus
cannot be continuously deformable into one another%
---they are topologically distinct.
As a matter of fact,
topology gives rise to new physics \cite{Ryder,Nakahara}
with some of the following relevant features:
\newcounter{counterTopologyFeatures}\setcounter{counterTopologyFeatures}{1}%
(\roman{counterTopologyFeatures})\stepcounter{counterTopologyFeatures}~%
metastable configurations
(mostly solitons)
fall into distinct classes,
of which the trivial configuration
(vacuum) belongs only to one class;
(\roman{counterTopologyFeatures})\stepcounter{counterTopologyFeatures}~%
there exists no superposition principle,
\textit{i.e.} resultant configurations form complicated ones;
(\roman{counterTopologyFeatures})\stepcounter{counterTopologyFeatures}~%
frustration phenomena may arise from topological obstruction
leading to both global and localized effects 
\cite{VSDTS,TSGF,HSETS,HSCS}.

N{\oe}ther-like calculational machineries exist
to study topological systems:
vesicles were shown \cite{BDVAG} to be subject
to a technique known as the Bogomol'nyi decomposition
which successfully treats various topological models
in fields ranging from condensed matter physics to high energy physics
\cite{Ryder,BogoE,BelavinPolyakov,Bogomolnyi,Felsager}.
Concisely, the Bogomol'nyi technique applies to
the total integral of the contracted self-product of the shape tensor,
known to be globally invariant
under conformal transformations \cite{AICM,OPCW},
revealing the topological nature of bending phenomena by both
identifying the non-trivial metastable bending configurations
and splitting bending configurations
into topological classes according to their genus/end.
The non-trivial metastable bending configurations are
the round sphere and the minimal surfaces
up to a conformal transformation of the ambient space
(\textit{e.g.} catenoid and Lawson surfaces \cite{CMSS3})
in agreement with observations
\cite{PLIPO,XHBP8,XAmMn,CFMV,ETSTV}.
Furthermore, it is worth noticing that
any deformation of the non-trivial metastable bending configurations
spontaneously matches,
for vesicles of spherical topology,
the deviatoric bending contribution described by Fischer
\cite{BSLBIII,BSLBV}
and, for vesicles of non-spherical topology,
the mean curvature bending contribution
(up to a conformal transformation of the ambient space)
described by Helfrich
\cite{XHBP8,XAmMn,CFMV,HelfrichZN2}.
For completeness, we note that
the involved total integral corresponds to
the covariant form of the bending energy
proposed by Canham \cite{Canham}.
Clearly, at least to study global bending phenomena,
the topological approach is very appealing.
Nevertheless, the mathematics involved here may discourage 
some readers.  To overcome this, we show how this unorthodox 
approach leads to a description of spherical vesicles of 
revolution in terms of `frustrated' sine-Gordon kinks.

\section{Self-dual approach}
Before focusing on spherical vesicles of revolution, we 
first succinctly expose how topology in vesicles is revealed
by the Bogomol'nyi decomposition
\cite{Ryder,BogoE,BelavinPolyakov,Bogomolnyi,Felsager}.
Following
the fundamental theorem of surface theory,
we represent the vesicle shape by a pair of tensors
coupled to each other by integrability conditions,
namely a prescribed first fundamental form
(the metric tensor) $g_{ij}$ coupled to a prescribed second 
fundamental form (the shape tensor) $b_{ij}$ with respect to
the equation of Gauss and 
the equations of Codazzi and Peterson
\cite{Struik,MDGCSM,GPI}.
For the bending Hamiltonian,
we consider the covariant functional
\begin{equation}
\label{BP/Hb/functional/def}
\stHmBnd\left[\stSurface\right]%
	\equiv%
  {\textstyle{\frac{1}{2}}}\stKb\!%
	{\int_{\stSurface}}\!%
  \stdS\;%
  b_{ij}b^{ij}%
  ,
\end{equation}
which depends on the vesicle shape $\stSurface$
through the prescribed pair $(g_{ij},b_{ij})$
and on the phenomenological parameter $\stKb$
describing the bending rigidity.
The summation convention has been adopted while
customary notations have been used:
the integral runs over the surface manifold $\stSurface$
with surface element $\stdS=\std{x^2}\sqrt{\left|g\right|}$,
where $\left|g\right|$ represents the determinant $\det(g_{ij})$
and $x$ the set of arbitrary intrinsic coordinates.
Observe that the suggested covariant functional (\ref{BP/Hb/functional/def})
corresponds to the covariant form of the bending energy invoked  
by Canham \cite{Canham} since the scalar $b_{ij}b^{ij}$ is equal to the 
trace of the square of the mixed shape tensor $b_{i}^{\hphantom{i}j}$,
and thus to the sum of its squared eigenvalues,
namely the sum of the squared principal curvatures.
Next, we define the dual tensor $\dual{a}_{ij}$
of an arbitrary tensor $a_{ij}$ by
\begin{equation}
\label{BP/duality/def}
\dual{a}_{ij}%
	\equiv%
	\stLevCS_{ik}\stLevCS_{jl}\:%
	a^{kl}%
\end{equation}
where $\stLevCS_{mn}$ denotes
the totally antisymmetric tensor
associated with the surface $\stSurface$.
It is easily checked that
the dual transformation ${a_{ij}}\to{\dual{a}_{ij}}$
is an involution,
\textit{i.e.}
\begin{equation}
\label{BP/duality/involution}
	\dual{\dual{a}}_{ij}%
	=%
	{a}_{ij}%
	.
\end{equation}
While there exists no such relation between
the shape tensor $b_{ij}$ and its dual $\dual{b}_{ij}$,
the metric tensor $g_{ij}$ satisfies
the property of self-duality;
that is 
\begin{equation}
\label{BP/metric/self-duality}
	\dual{g}_{ij}=g_{ij}%
	.
\end{equation}
Nevertheless
the bending Hamiltonian (\ref{BP/Hb/functional/def})
remains locally invariant under the dual transformation
as we have, after straightforward index rearrangements,
\begin{equation}
\label{BP/Hb/density/duality}
	\dual{b}_{ij}\dual{b}^{ij}%
	=%
	b_{ij}b^{ij}%
	.
\end{equation}
Most significantly,
the extrinsic curvature $\stEC$,
which yields
\begin{equation}
\label{BP/Hb/density/curvature}
	\stEC%
	=%
	{\textstyle{\frac{1}{2}}}\:%
	\dual{b}_{ij}%
	b^{ij}
	,
\end{equation}
remains clearly unchanged under the dual transformation
according to the involutive relation (\ref{BP/duality/involution}).
The interesting feature is that
the extrinsic curvature $\stEC$
coincides with the intrinsic curvature 
(the Gaussian curvature)
$\stGC$ when the embedding ambient space is flat:
its total integral then measures 
the topology of the vesicle shape.

More precisely,
the Gauss-Bonnet theorem \cite{Struik,CEMSFTC}
claims that
\begin{equation}
\label{BP/GaussBonnetChern/flat}
{\int_{\stSurface_{\stgenus,\stends}}}\!\!\!\!%
  \stdS\:%
  \stEC%
  =%
  -4\pi%
	\left(\stgenus\!+\!\stends\!-\!1\right)%
   , 
\end{equation}
where $\stSurface_{\stgenus,\stends}$ is a surface manifold
embedded in a flat ambient space and
topologically equivalent to a closed surface manifold
of genus $\stgenus$ less $\stends$ points (ends).
These considerations lead to
the construction of a self-dual theory
by introducing the self-dual/anti-self-dual tensors
\begin{equation}
\label{BP/Hb/tensor/deviatoric/def}
{B^{\pm}}_{ij}%
	\equiv%
	{\textstyle{\frac{1}{\sqrt{2}}}}%
	\left[
		{b}_{ij}%
		\pm%
		\dual{b}_{ij}%
	\right]
	,
\end{equation}
which manifestly verify
the property of self-duality/anti-self-duality
according to the involutive relation (\ref{BP/duality/involution}):
\begin{equation}
\label{BP/Hb/tensor/deviatoric/self-duality}
	\dual{B^{\pm}}_{ij}%
	=%
	\pm\:%
	{B^{\pm}}_{ij}%
	.
\end{equation}
Contracting the self-product of the self-dual/anti-self-dual tensors (\ref{BP/Hb/tensor/deviatoric/def})
and using the equalities (\ref{BP/Hb/density/duality}) and (\ref{BP/Hb/density/curvature})
readily show that the bending Hamiltonian density $\stHmBndD$
extracted from the bending Hamiltonian (\ref{BP/Hb/functional/def})
decomposes as
\begin{equation}
\label{BP/Hb/Bogo/decomposition/density}
\stHmBndD%
	=%
	{\textstyle{\frac{1}{2}}}\stKb\:%
	{B^{\pm}}_{ij}{B^{\pm}}^{ij}%
	\;\mp\;%
	\stKb\:%
	\stEC%
	.
\end{equation}
Integrating this decomposition
(\ref{BP/Hb/Bogo/decomposition/density})
over the surface manifold $\stSurface$ and
recognizing the total curvature (\ref{BP/GaussBonnetChern/flat})
leads to rewriting the bending Hamiltonian (\ref{BP/Hb/functional/def})
in each topological class specified by the pair $(\stgenus,\stends)$
in the form
\begin{equation}
\label{BP/Hb/Bogo/decomposition/Hamiltonian}
\stHmBnd\left[\stSurface_{\stgenus,\stends}\right]%
	=%
	{\textstyle{\frac{1}{2}}}\stKb\!%
	{\int_{\stSurface_{\stgenus,\stends}}}\!\!\!\!\!%
	\stdS\;%
	{B^{\pm}}_{ij}{B^{\pm}}^{ij}%
	\pm
	4\pi\stKb%
	\left(\stgenus\!+\!\stends\!-\!1\right)%
	.
\end{equation}
Since the tensors ${B^{\pm}}_{ij}$ yield
the precious inequalities
\begin{equation}
\label{BP/Hb/tensor/deviatoric/inequalities}
	{B^{\pm}}_{ij}{B^{\pm}}^{ij}%
	\geqslant0
	,
\end{equation}
the decompositions (\ref{BP/Hb/Bogo/decomposition/Hamiltonian})
saturate when the tensors ${B^{\pm}}_{ij}$ vanish.
Henceforth,
since the shape tensor is self-dual
(${B^{-}}_{ij}\!=\!0$)
only for the round sphere ${\mathbb{S}}^2$,
the decomposition with sign $(-)$
in (\ref{BP/Hb/Bogo/decomposition/Hamiltonian})
is relevant only for the vesicle surfaces $\stSurface_{0}$
topologically equivalent to the round sphere ${\mathbb{S}}^2$;
we get  
\begin{equation}
\label{BP/Hb/Bogo/decomposition/Hamiltonian/spherical}
\stHmBnd\left[\stSurface_{0}\right]%
	=%
	{\textstyle{\frac{1}{2}}}\stKb\!%
	{\int_{\stSurface_{0}}}\!\!\!%
	\stdS\;%
	{B^{-}}_{ij}{B^{-}}^{ij}%
	+%
	4\pi\stKb%
	.
\end{equation}
Similarly,
as the shape tensor is anti-self-dual
(${B^{+}}_{ij}\!=\!0$)
only when the surface manifold is a minimal surface,
the decomposition with sign $(+)$
in (\ref{BP/Hb/Bogo/decomposition/Hamiltonian})
is pertinent only for
the vesicle surfaces $\stSurface_{\stgenus,\stends}$
topologically equivalent to
a minimal surface of genus $\stgenus$ with $\stends$ ends.
Since within flat space
there is no closed minimal surface $(\stends\!=\!0)$ whereas
minimal surfaces of genus $\stgenus$ $(\stgenus\!\geqslant\!0)$
with $\stends$ ends $(\stends\!\geqslant\!2)$ do exist \cite{CEMSFTC},
we write
\begin{equation}
\label{BP/Hb/Bogo/decomposition/Hamiltonian/non-spherical}
\stHmBnd\left[\stSurface_{\stgenus,\stends}\right]%
	=%
	{\textstyle{\frac{1}{2}}}\stKb\!%
	{\int_{\stSurface_{\stgenus,\stends}}}\!\!\!\!\!%
	\stdS\;%
	{B^{+}}_{ij}{B^{+}}^{ij}%
	+%
	4\pi\stKb%
	\left(\stgenus\!+\!\stends\!-\!1\right)%
\end{equation}
which is valid only
when $\stgenus\geqslant0$ and $\stends\geqslant2$.
Notice that the minimal surface of genus zero with two ends is the catenoid.
Tedious considerations allow one
to expand the previous decomposition
(\ref{BP/Hb/Bogo/decomposition/Hamiltonian/non-spherical})
to closed surfaces of arbitrary genus $(\stgenus\geqslant1)$
\cite{BDVAG}.
Besides bringing out the inherent topology,
the Bogomol'nyi decompositions 
(\ref{BP/Hb/Bogo/decomposition/Hamiltonian/spherical})
and (\ref{BP/Hb/Bogo/decomposition/Hamiltonian/non-spherical})
offer a new perspective on bending phenomena:
when the shape tensor
obeys the local property of self-duality/anti-self-duality
then the bending Hamiltonian is saturated. 
Conversely, when the shape tensor
violates the local property of self-duality/anti-self-duality
then the bending Hamiltonian is frustrated,
\textit{i.e.}
an extra energy contribution is spontaneously created
\cite{BDVAG,VSDTS,TSGF,HSETS,HSCS}
that tends to vanish globally in the system. 
Before following this matter any further,
we turn our attention to
the spherical vesicles of revolution
which are of pertinent interest.

\section{Spherical vesicles of revolution}
Now, we concentrate on vesicle shapes that are both 
smoothly transformable to the round sphere ${\mathbb{S}}^2$
and can be generated by rotating a two-dimensional curve
(a profile) $\stProfile$
about an axis,
\textit{i.e.}
on spherical vesicles of revolution.
For our purpose,
the azimuthal isometric coordinates $(u,\varphi)$
appear to be a suitable choice of coordinates as follows.
First, adopting isometric coordinates (`isothermal' coordinates)
allows us to write the metric tensor $g_{ij}$ in the form
\cite{Struik,Nakahara,TMTBH}
\begin{equation}
\label{BP/SVR/metric}
	g_{uu}=g_{\varphi\varphi}=\stE^{2\stWeylScalingExp(u)}%
\end{equation}
where the local Weyl gauge field $\stWeylScalingExp$
\cite{Fulton1962}
depends only on the isometric coordinate $u$
because of the azimuthal symmetry.
Moreover, for any surface of revolution,
analytical manipulations  
show that the shape tensor $b_{ij}$ 
in azimuthal isometric coordinates $(u,\varphi)$ yields
\begin{equation}
\label{BP/SVR/shape}
	b_{uu}\!=%
		\!-\stE^{\stWeylScalingExp(u)}
		\partial_{u}\stNPolarAngle(u)%
	\quad{and}\quad%
	b_{\varphi\varphi}\!=%
		\!-\stE^{\stWeylScalingExp(u)}
		\sin\stNPolarAngle(u)%
        , 
\end{equation}
where $\stNPolarAngle$ corresponds to the polar angle
of the unit normal vector
of the surface.
However,
in order for the two tensors $g_{ij}$ and $b_{ij}$,
cast respectively 
in the forms (\ref{BP/SVR/metric}) and (\ref{BP/SVR/shape}),
to be the first and second fundamental forms
for a surface $\stSurface$ in ${\mathbb{R}}^3$,
they must satisfy 
the Gauss equation and 
the Codazzi and Peterson equations
\cite{Struik,MDGCSM,GPI}:
in this case,
after easy computations, 
this set of integrability conditions
reduces to the equation
\begin{equation}
\label{BP/SVR/intcond}
	\partial_{u}\stWeylScalingExp(u)%
	=%
		\cos\stNPolarAngle(u)%
	.
\end{equation}
Conversely, simple calculations show that
the Gauss-Weingarten equations
\cite{Struik,MDGCSM,GPI}
associated to the first form (\ref{BP/SVR/metric})
and the second form (\ref{BP/SVR/shape})
under the integrability condition (\ref{BP/SVR/intcond})
give a unique surface of revolution,
except for both its position and scale in space.
To summarize,
any pair of tensors $(g_{ij},b_{ij})$ that verifies
the formulae
(\ref{BP/SVR/metric}), (\ref{BP/SVR/shape}) and (\ref{BP/SVR/intcond})
in azimuthal isometric coordinates $(u,\varphi)$
is equivalent
to a surface of revolution $\stSurfaceOfRevolution$ in ${\mathbb{R}}^3$.
Substituting (\ref{BP/SVR/metric}) and (\ref{BP/SVR/shape})
into (\ref{BP/Hb/tensor/deviatoric/def}) leads to
\begin{equation}
\label{BP/SVR/deviatoric}
	{B^{\pm}}_{uu}=\pm{B^{\pm}}_{\varphi\varphi}%
		=%
		{\textstyle{\frac{-1}{\sqrt{2}}}}\,%
		\stE^{\stWeylScalingExp(u)}%
		\bigl[
			\partial_{u}\stNPolarAngle(u)%
			\pm%
			\sin\stNPolarAngle(u)%
		\bigr]
	.
\end{equation}
Henceforth,
by applying (\ref{BP/Hb/Bogo/decomposition/Hamiltonian/spherical}),
the bending Hamiltonian Bogomol'nyi decomposition
for the spherical surfaces of revolution $\stSurfaceOfRevolution_{0}$
reads
\begin{equation}
\label{BP/SVR/Hb/Bogo/decomposition/Hamiltonian}
\stHmBnd\left[\stSurfaceOfRevolution_{0}\right]%
	=%
	\pi\stKb\!%
	{\int_{\stProfile}}\!%
	\std{u}\;%
	\bigl[
		\partial_{u}\stNPolarAngle(u)%
		-%
		\sin\stNPolarAngle(u)%
	\bigr]^2%
	+%
	4\pi\stKb%
        , 
\end{equation}
where the integral runs along the profile $\stProfile$
of the spherical surface $\stSurfaceOfRevolution_{0}$.
The bending Hamiltonian (\ref{BP/SVR/Hb/Bogo/decomposition/Hamiltonian})
is saturated when the polar angle $\stNPolarAngle$
satisfies the equation
\begin{equation}
\label{BP/SVR/Hb/saturation/equation}
	\partial_{u}\stNPolarAngle(u)=\sin\stNPolarAngle(u)%
	,
\end{equation}
whose centered solution is the sine-Gordon kink
\begin{equation}
\label{BP/SVR/Hb/saturation/SG}
	\stNPolarAngle(u)=%
		\pi-\arccos(\tanh(u))%
	.
\end{equation}
Resolving the Gauss-Weingarten equations
\cite{Struik,MDGCSM,GPI}
with respect to
(\ref{BP/SVR/metric}), (\ref{BP/SVR/shape}), (\ref{BP/SVR/intcond})
and (\ref{BP/SVR/Hb/saturation/SG})
uniquely gives
(except for its position and scale in space)
the surface, in cylindrical parametrization,
\begin{equation}
\label{BP/SVR/Hb/saturation/surface}
	r(u)=%
		\;
		\sech(u)
	,
	\qquad%
	z(u)=%
		-\tanh(u)
	,
\end{equation}
which is the round sphere ${\mathbb{S}}^2$
in azimuthal isometric coordinates $(u,\varphi)$,
$u$ running from $-\infty$ to $+\infty$.
Hence,
in accordance with the existence theorem previously invoked,
the spherical vesicle of revolution which saturates
the bending Hamiltonian (\ref{BP/Hb/functional/def})
is the round sphere ${\mathbb{S}}^2$.
However the key result is contained in the fact that
the metastable spherical vesicle of revolution
can be described by a saturated sine-Gordon kink:
any spherical vesicle of revolution
may be thereby envisioned as a frustrated
(\textit{i.e.} unsaturated) sine-Gordon kink,
and highly nonlinear global physics
may henceforth emerge from the model
\cite{VSDTS,TSGF,HSETS,HSCS}.
Therefore, without loss of generality,
the initial bidimensional system has been reduced to a well 
known unidimensional nonlinear system.
Next we illustrate our approach through an 
illuminating example.

\section{Latex bead on a spherical vesicle} 
\begin{figure}[t]
	\begin{center}
		\includegraphics[width=.7071\linewidth]{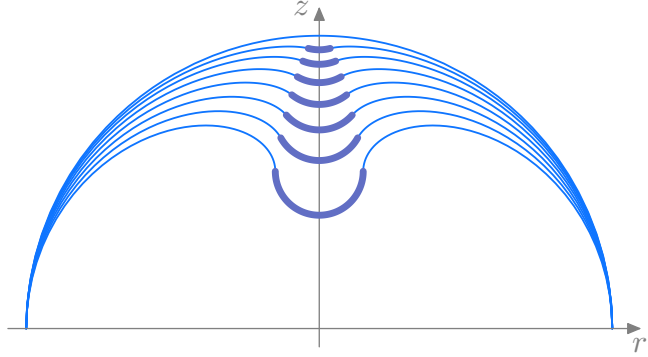}%
	\end{center}
	\caption{
		North hemispheric profiles
		of a spherical vesicle of revolution
		distorted by two identical round latex beads
		grafted at its poles
		with respect to different encapsulated radii $\stBeadEncapsulatedRadius$
		as deduced by the self-dual bending theory.
		The bold arcs correspond to the profiles
		of the polar cap imposed by the bead.
		Dimensionless latex bead radius $\stBeadRadius\!=\!0.15$;
		relative encapsulated radius ${\stBeadEncapsulatedRadius}/{\stBeadRadius}$
		from outside to inside:
		$0$,
		$\frac{1}{4}$,
		$\frac{3}{8}$,
		$\frac{1}{2}$,
		$\frac{5}{8}$,
		$\frac{3}{4}$,
		$\frac{7}{8}$,
		$1$.
		}%
\label{fig/NorthHemisphericProfiles}
\end{figure}

\begin{figure}[t]
	\begin{center}
		\includegraphics[width=.7071\linewidth]{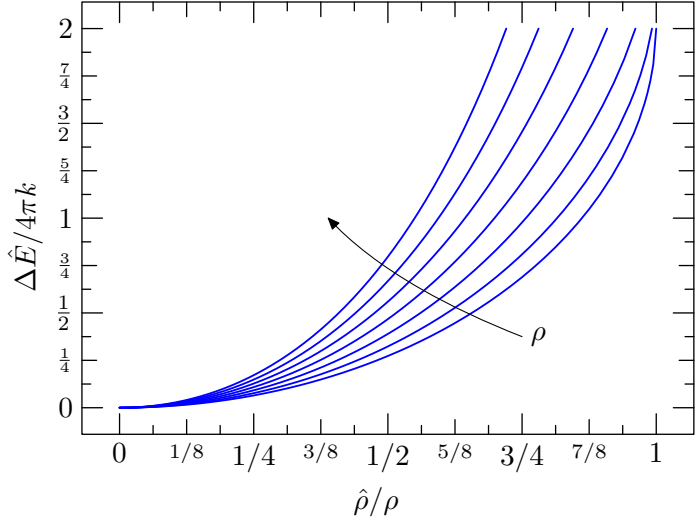}%
	\end{center}
	\caption{
		The relative extra bending energy ${\stExtraBendingEnergy}/{4\pi\stKb}$
		\textit{versus}
		the relative encapsulated radius ${\stBeadEncapsulatedRadius}/{\stBeadRadius}$
		for different radii $\stBeadRadius$ of the grafted latex beads:
		the radius $\stBeadRadius$ increases
		in the direction of the arrow.
		Dimensionless latex bead radius from right to left:
		$0.01$,
		$0.1$,
		$0.2$,
		$0.3$,
		$0.4$,
		$0.5$,
		$0.6$.
		}%
\label{fig/ExtraEnergyPlots}
\end{figure}

Grafting a latex bead on a spherical vesicle
induces a global distortion characterized by
two polar concave regions:
a pinched concave region around the bead faces a smooth one
as if a second bead were diametrically opposite
\cite{Koltover1999}.
Accordingly, we assume a bare spherical vesicle of revolution
with two identical round latex beads
which are respectively attached at the north and south poles.
\begin{figure}[b]
	\begin{center}
		\includegraphics[width=.7071\linewidth]{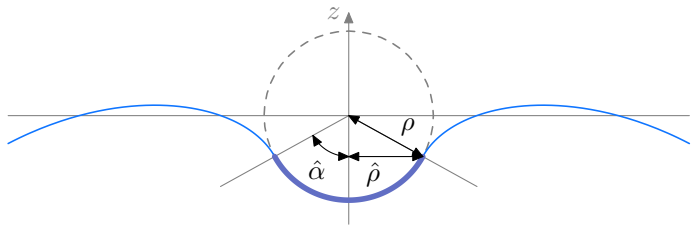}%
	\end{center}
	\caption{
		North pole sketch
		of a spherical vesicle of revolution
		distorted by two identical round latex beads
		grafted at its poles
		as deduced by the self-dual bending theory.
		The dashed circle profiles the bead,
		the bold arc the polar cap imposed by it.
		The arrows indicate the nomenclature:
		the latex bead dimensionless radius $\stBeadRadius$,
		the encapsulated dimensionless radius $\stBeadEncapsulatedRadius$,
		and the encapsulation angle $\stEncapsulationAngle$.
		Dimensionless latex bead radius $\stBeadRadius\!=\!0.15$;
		relative encapsulated radius
		${\stBeadEncapsulatedRadius}/{\stBeadRadius}\!=\!\frac{7}{8}$.
		}%
\label{fig/NorthPoleSketch}
\end{figure}
By bare, we mean that only
the bending Hamiltonian (\ref{BP/Hb/functional/def}) is considered. 
To begin with,
substituting (\ref{BP/SVR/metric}) and (\ref{BP/SVR/shape})
into (\ref{BP/Hb/functional/def}) gives
the Hamiltonian of our system in the canonical form
\begin{equation}
\label{BP/SVR/Hb/canonic/Hamiltonian}
\stHmBnd\left[\stSurfaceOfRevolution\right]%
	=%
	\pi\stKb\!%
	{\int_{\stProfile}}\!%
	\std{u}\;%
	\left[
		{\partial_{u}\stNPolarAngle(u)}^{2}%
		+%
		{\sin^{2}\stNPolarAngle(u)}%
	\right]%
	.
\end{equation}
The Euler-Lagrange equation
derived from (\ref{BP/SVR/Hb/canonic/Hamiltonian})
(or from (\ref{BP/SVR/Hb/Bogo/decomposition/Hamiltonian}))
is the sine-Gordon equation
\begin{equation}
\label{BP/SVR/Hb/SG/equation}
	\partial_{uu}\stNPolarAngle(u)%
	=%
		\sin\stNPolarAngle(u)\,%
		\cos\stNPolarAngle(u)%
	.
\end{equation}
The solution of (\ref{BP/SVR/Hb/SG/equation})
extending (\ref{BP/SVR/Hb/saturation/SG}),
\begin{equation}
\label{BP/SVR/Hb/SG/kink}
	\stNPolarAngle(u)=%
		\pi-\arccos(\JacobiSN(\textstyle{\frac{u}{\sqrt{\Jacobim}}}\!\mid\!\Jacobim))%
	\quad%
	\Jacobim\in\left(0,1\right]%
	,
\end{equation}
corresponds to the surface of revolution
\begin{equation}
\label{BP/SVR/Hb/SG/surface/parametrization}
	\eqalign{%
		r(u)=%
		&
			{\textstyle{\frac{1}{1+\sqrt{\Jacobim}}}}%
			\left[%
				\JacobiDN(\textstyle{\frac{u}{\sqrt{\Jacobim}}}\!\mid\!\Jacobim)%
				+\sqrt{m}%
				\JacobiCN(\textstyle{\frac{u}{\sqrt{\Jacobim}}}\!\mid\!\Jacobim)%
			\right]%
		,
		\\ 
		z(u)=%
		&
			{\textstyle{\frac{-1}{1+\sqrt{\Jacobim}}}}%
			\left[%
				\sqrt{m}%
				\JacobiSN(\textstyle{\frac{u}{\sqrt{\Jacobim}}}\!\mid\!\Jacobim)%
				+%
				\JacobiEpsilon(\textstyle{\frac{u}{\sqrt{\Jacobim}}}\!\mid\!\Jacobim)%
				-%
				{\textstyle{\frac{(1-\Jacobim)u}{\sqrt{\Jacobim}}}}%
			\right]%
	,
	}
\end{equation}
which smoothly joins%
	\footnote[1]{%
	The junction conditions are the ones implicitly assumed
	in Ref.~\cite{Koltover1999}:
	at the junction points,
	the normal lines of the vesicle surface and of the latex bead surface
	are imposed to coincide%
	---see \figurename~\ref{fig/NorthPoleSketch}.
	Such an assumption is reasonable at the vesicle scale,
	nevertheless at the membrane scale an adequate treatment
	may be needed since the curvature experiences a discontinuity
	at the junction points%
	---as far as we know there is no such treatment in the literature.
	}
the concavely bounded spherical polar caps to the vesicle surface 
when the parameter $\Jacobim$ satisfies
\begin{equation}
\label{BP/SVR/Hb/SG/surface/parameter}
	\Jacobim=%
		\left[%
			\frac{\stBeadRadius\,(1-\stBeadEncapsulatedRadius^{2})}%
				{\stBeadRadius\,(1+\stBeadEncapsulatedRadius^{2})+2\stBeadEncapsulatedRadius^{2}}%
		\right]^{2}%
	\qquad%
	0\leqslant\stBeadEncapsulatedRadius\leqslant\stBeadRadius<1\;%
  , 
\end{equation}
$\stBeadRadius$ and $\stBeadEncapsulatedRadius$ being respectively
the latex bead and encapsulated dimensionless radii%
---see \figurename~\ref{fig/NorthPoleSketch} and \figurename~\ref{fig/NorthHemisphericProfiles}.
The isometric coordinate $u$ runs
from $-\hat{u}$ to $+\hat{u}$ with
	$%
		\hat{u}\!=\!%
			 \sqrt{\Jacobim}[2\EllipticK(\Jacobim)%
			-\EllipticF(\cos\stEncapsulationAngle\!\mid\!\Jacobim)]%
	$,
where the encapsulation angle $\stEncapsulationAngle$ yields
$\sin\stEncapsulationAngle\!=\!\stBeadEncapsulatedRadius/\stBeadRadius$.
The reader is encouraged to check that,
when the encapsulated dimensionless radius $\stBeadEncapsulatedRadius$ vanishes, 
the parameter $\Jacobim$ effectively tends towards $1$ and thus  
the surface (\ref{BP/SVR/Hb/SG/surface/parametrization}) tends to
the round sphere ${\mathbb{S}}^2$ (\ref{BP/SVR/Hb/saturation/surface})
as expected.
Now the extra energy $\stExtraBendingEnergy$ spontaneously generated
by the polar latex beads can be computed.
We have
\begin{equation}
\label{BP/SVR/Hb/SG/deviatoric}
	{B^{-}}_{uu}=-{B^{-}}_{\varphi\varphi}%
		=%
		{\textstyle{\frac{-1}{\sqrt{2}}}}\,%
		{\textstyle{\frac{1-\sqrt{\Jacobim}}{\sqrt{\Jacobim}}}}%
                , 
\end{equation}
from which we obtain
\begin{equation}
\label{BP/SVR/Hb/SG/extra}
\fl%
	\stExtraBendingEnergy=%
		\frac{4\pi\stKb}{\sqrt{\Jacobim}}%
		\biggl[%
			2%
			\left[%
				\EllipticE(\Jacobim)%
				-%
				{\textstyle{\frac{1-\Jacobim}{2}}}%
				\EllipticK(\Jacobim)%
			\right]%
			-%
			\left[%
				\EllipticE(\cos\stEncapsulationAngle\!\mid\!\Jacobim)%
				-%
				{\textstyle{\frac{1-\Jacobim}{2}}}%
				\EllipticF(\cos\stEncapsulationAngle\!\mid\!\Jacobim)%
			\right]%
			-%
			\sqrt{\Jacobim}%
			\cos\stEncapsulationAngle%
		\biggr]%
		.
\end{equation}
Standard notations have been adopted%
	\footnote[2]{%
		The functions $\JacobiSN$, $\JacobiCN$ and $\JacobiDN$ are
		the Jacobian elliptic functions,
		$\EllipticF$ and $\EllipticK$ 
		the incomplete and complete elliptic integrals of the first kind,
		$\EllipticE$
		the incomplete and complete
		(depending on the argument)
		elliptic integrals of the second kind,
		and the function $\JacobiEpsilon$ denotes
		the Jacobian Epsilon function defined as
		$%
			\JacobiEpsilon(u\!\mid\!\Jacobim)\!\equiv\!%
				{\int_{0}^{u}}\std{t}\;{\JacobiDN^{2}}(t\!\mid\!\Jacobim)%
		$
		\cite{Abramowitz}.
	}%
\cite{Abramowitz,Whittaker,ByrdFriedman}.
Observing that,
according to (\ref{BP/SVR/Hb/SG/surface/parameter}),
the parameter $\Jacobim$ decreases strictly from $1$ to $0$
with respect to the encapsulated radius $\stBeadEncapsulatedRadius$,
allows an effortless description:
when the encapsulated radius $\stBeadEncapsulatedRadius$ vanishes,
the underlying soliton (\ref{BP/SVR/Hb/SG/kink}) 
reaches its saturated configuration (\ref{BP/SVR/Hb/saturation/SG});
when the encapsulated radius $\stBeadEncapsulatedRadius$ increases,
the underlying soliton (\ref{BP/SVR/Hb/SG/kink}) becomes frustrated
and the bending energy is accordingly altered%
---see \figurename~\ref{fig/ExtraEnergyPlots}.
Finally,
our naive model system demonstrates
at least one weakness of the traditional (local) approach
for the description of global bending phenomena, as follows:
since its mean curvature $\stMC$,
or the half trace of the shape tensor,
is constant,
\begin{equation}
	\stMC%
		=%
		{\textstyle{\frac{1}{2}}}\,%
		{\textstyle{\frac{1+\sqrt{\Jacobim}}{\sqrt{\Jacobim}}}}%
	,
\end{equation}
the surface of revolution
(\ref{BP/SVR/Hb/SG/surface/parametrization})
causes the harmonic Helfrich bending Hamiltonian density
either to vanish if a phenomenological spontaneous curvature is introduced
or, if not, to be proportional to the square of the mean curvature
\cite{XHBP8,XAmMn,CFMV,HelfrichZN2},
irrespective of the value of the encapsulated radius $\stBeadEncapsulatedRadius$.
Consequently the traditional approach is partially insensitive  
to the encapsulation mechanism experienced by our system
in the sense that an observer
somewhere on the surface outside the grafted regions
cannot determine whether or not latex beads are grafted at the poles
just from the harmonic Helfrich bending Hamiltonian density,
whereas the same observer can obtain a quantitative answer
by computing the extra bending Hamiltonian density.
In the former case, what occurs globally is not known 
because in every scenario
the harmonic Helfrich bending Hamiltonian density
takes the same form;
in the latter,
the extra bending Hamiltonian density vanishes
when there are no latex beads
but it measures the deformation
otherwise.
Nevertheless it may be objected that,
if no phenomenological spontaneous curvature is introduced,
computing the harmonic Helfrich bending Hamiltonian density
allows for an answer which is quantitatively similar,
provided that the expression is known for the particular state,
the one without latex beads for example.
Such exact information, required for the former approach but
superfluous for the topological approach,
may not be so easily identified in a more realistic case.
Second, the topological approach provides actually a finer response
as the deformation measurement is really enabled 
by the anti-self-dual tensor
which can be conceived as a bending deformation tensor.

\section{Conclusion} 
In conclusion,
the results of our investigations not only provide 
a novel viewpoint regarding bending phenomena,
but also demonstrate the possibility of describing
spherical vesicles of revolution
by frustrated sine-Gordon kinks
as a method to investigate the essence of
highly nonlinear global behavior observed in vesicles.
The geometry of vesicles has been studied
quite extensively in recent years
\cite{MESPFMV,CapovillaGuven2002,LenzNelson2002,LinLoTsai2003}
but it was not considered in conjunction
with the global topological aspects.
Of course, the present model is too naive to capture
the complexity of biological vesicles, in particular
it should be complemented by the inclusion of material fields and 
constraints
\cite{XHBP8,XAmMn,CFMV}.
Still, as far as bending phenomena are concerned,
the self-dual bending theory should provide
a powerful theoretical framework
to study global behavior observed in vesicles or, in 
general, other deformable systems.

\ack 
This work was supported in part by the U.S. Department of 
Energy and in part by
E.C. contract HPRN-CR-1999-00163 (LOCNET network)
and in part by
the Thuringian Ministry of Science, Research and Art, Germany.

\section*{References}
\bibliographystyle{ioppnum}
\bibliography{sdbtv}

\end{document}